\documentstyle[12pt]{article}
\begin{document}
\def\thebibliography#1{\section*{REFERENCES\markboth
 {REFERENCES}{REFERENCES}}\list
 {[\arabic{enumi}]}{\settowidth\labelwidth{[#1]}\leftmargin\labelwidth
 \advance\leftmargin\labelsep
 \usecounter{enumi}}
 \def\newblock{\hskip .11em plus .33em minus -.07em}
 \sloppy
 \sfcode`\.=1000\relax}
\let\endthebibliography=\endlist
\hoffset = -1truecm
\voffset = -2truecm
\title{\large\bf
On Simulating Liouvillian Flow From Quantum Mechanics Via Wigner Functions  
}
\author{
{\normalsize\bf
A.N.Mitra \thanks{e.mail: (1) insa@giasdl01.vsnl.net.in(subj:a.n.mitra);
(2) anmitra@csec.ernet.in}
}\\
{\normalsize 244 Tagore Park, Delhi-110009, India.} \and
{\normalsize\bf
R.Ramanathan
}\\
\normalsize Dept. of Physics, Univ. of Delhi, Delhi-110007, India.}
\date{13th November 1997}
\maketitle
\begin{abstract}
The interconnection between quantum mechanics and probabilistic classical
mechanics for a free relativistic particle is derived in terms of Wigner 
functions (WF) for both Dirac and Klein-Gordon (K-G) equations. Construction 
of WF is achieved by first defining a bilocal 4-current and then taking its 
Fourier transform w.r.t. the relative 4-coordinate. The K-G and Proca cases 
also lend themselves to a closely parallel treatment {\it provided} the 
Kemmer-Duffin $\beta$-matrix formalism is employed for the former. 
Calculation of WF is carried out in a Lorentz-covariant fashion by standard 
`trace' techniques. The results are compared with a recent derivation due 
to Bosanac. \\\\      
PACS: 05.60.+w ; 03.65.Ca ; 11.10.Qr ; 11.40.-q
\end{abstract} 
  
\newpage


\section{Introduction: Wigner Function As Link Between QM and CM}

Since the advent of Quantum Mechanics (QM), there have been several 
investigations aimed at understanding Schroedinger equations and their 
relativistic counterparts from stochastic /statistical premises in classical 
mechanics, epitomised by Focker-Planck [1,2] and  Liouville equations [3-6]. 
In the process most such investigations have had to resort at some stage to 
some sort of ansatz designed to introduce  the `square-root' of a probability 
function [1-4], or make a more direct use of the wave function from the outset
[5,6]. This of course does not address the core issue of the foundations
of quantum physics, viz., the existence of complex probability amplitudes
obeying the ``superposition principle'', and the need to invoke the ``collapse
hypothesis'' to explain the outcome of a quantum measurement. Even without any
commitment to resolve this central mystery of conventional quantum physics,
it is both interesting and illuminating to explore the classical roots of
quantum mechanics by comparing and relating their respective mathematical
structures as in several contemporary investigations [3-6].       
\par	
	Among the several approaches to an alternative `derivation' of QM,
we would like to pick out for explicit scrutiny, a recent formulation due 
to Bosanac [4] in which the author makes essential use of the probabilistic 
(instead of `trajectory-oriented') approach to Classical Mechanics (CM) via 
the Liouville equation, wherein  the Uncertainty Principle is sought to
be incorporated at the outset, as a means of capturing a good chunk of QM. 
Our purpose in this regard is more in the nature of an `inverse' (to [4])
approach, which stems from a desire to give a more direct role to the QM 
wave function born out of its `being already there', thereby according it 
a more open role in the construction of the Wigner Function [7], as in other
orthodox approaches [5,6]. As we shall see, this approach has the advantage 
of presenting the treatment of different types of relativistic equations 
(Dirac, Klein-Gordon, Proca, etc) within a unified framework.  
\par
	The central quantity that satisfies the Liouville equation is the 
Wigner-function [7] which in itself was conceived within the QM framework,
as a bilinear in the `wave function'. Granting the existence of the wave 
function in turn implies the existence of its Fourier transform which 
represents its momentum space counterpart, and the two together guarantee a 
built-in effect of the Uncertainty Principle, as clearly recognized in [4].  
Using such a (hybrid) approach in which the prior existence of the wave 
function is implicit in the structure of the Wigner function [7], Bosanac 
[4] claims to derive the Schroedinger equation from the Liouville equation, 
consistently with current conservation, apparently as an inverse(?) problem.  
An extension to the corresponding relativistic problem, rather surprisingly, 
seemed to yield only the Dirac equation (spinor), but not the Klein-Gordon 
(K-G) equation (scalar), a result whose mystery we seek to clarify in this 
paper and (hopefully) pave the way to a generalization (see below).
\par
	Specifically, we have been actuated by a desire to re-examine the 
issue of Dirac versus K-G [4],  but with a more conscious recognition of 
an active role of the `wave function' at the outset, and in the process we  
find that the Liouville and current conservation equations for both the Dirac 
and K-G cases  follow from identical premises, {\it provided} one has no 
qualms about the free use of {\it {multi-component}} wave functions in both 
Dirac and K-G cases, and not (selectively) for the Dirac case only [4]. 
Indeed it is easily seen from a detailed perusal of the steps in [4] that 
the author's success in deriving the Dirac equation is crucial to his 
{\it ansatz} of a 4-component spinor, without which the necessary consistency 
checks would not be forth-coming. On the other hand he did not seem to 
consider a similar possibility for the K-G case, viz., the existence of the 
less familiar Kemmer-Duffin equations for integral spins [8], wherein the 
multicomponent structure of the corresponding wave function satisfying a 
first order equation as in the Dirac case seems to be the right form for
putting the K-G and Proca cases on parallel footing with Dirac. Of course
the question of whether, inspite of a multicomponent ansatz for the K-G 
wave function, a bosanac-like treatment [4] of working backwards from the
Liouville-cum-current-conservation equations to the K-G equation would have
succeeded, is still open, since the Kemmer-Duffin $\beta$-matrices are 
singular, unlike the Dirac matrices which are not (see Sec.4). 
\par
	In carrying out this exercise we have been guided solely by the 
orthodox (Dyson [9]) philosophy that a relativistic quantization of a 
{\it single} particle has inherent problems of consistency, whose resolution 
hinges on the simultaneous existence of an infinite number of particles, 
i.e.,  on a {\it field} concept. Indeed, Dirac's postulation of the infinite 
sea (of negative energy particles) in the vacuum was enough ground for 
Pauli-Weisskopf [10] to reinterpret the single particle density from the K-G 
equation as an average charge density (which could `locally' have either 
sign) [9]. Of course from a practical point of view, a single particle 
interpretation has considerable appeal, but it requires the `protective 
cover' of Feynman's positron theory [11] which mandates $\pm$ time 
directions to be associated with $\pm$ energy propagations respectively.             
\par
	In what follows, we shall keep in the forefront, the results of [4]
for pointwise comparison with the derivation to be presented here under the
orthodox `field' paradigm [9] which is routinely available in any standard 
text [12].  In Sec.2, we give a `direct' derivation of the Wigner function 
for the Dirac equation and thence the corresponding Liouville plus current 
conservation equations, a procedure whose steps are  roughly in `inverse' 
order to those of [4]. In Sec.3, we do the corresponding exercise for 
the K-G and Proca cases via the Kemmer-Duffin [8] formulation [13], and in 
so doing, bring out the strong similarity of the two systems. Sec.4 
concludes with a short discussion, including a comparison with [4].     			
	 
\section{Wigner Function for the Dirac Equation}

We begin by writing the Dirac equation for a free field $\psi(x)$ in the 
`Euclidean' covariant notation as [13] 
\setcounter{equation}{0}
\renewcommand{\theequation}{2.\arabic{equation}}
\begin{equation}
({\gamma}_\mu \partial_\mu + m) \psi(x) = 0; \quad
{\bar \psi}(x) (m-\partial_\mu \gamma_\mu) = 0.
\end{equation}
Now define a bilocal current $j_\mu(x_1,x_2)$ as
\begin{equation}
j_\mu(x_1,x_2) = {\bar \psi}(x_1) i \gamma_\mu \psi(x_2)
\end{equation}
Next put $x_1 = x - q$ and $x_2 = x + q$ as in [4], and define a 4-vector
$V_\mu(x,p)$ as
\begin{equation}
V_\mu(x,p) = {\pi}^{-4} \int d^4q e^{2ip.q}  
{\bar \psi}(x+q){i\gamma_\mu} \psi(x-q).
\end{equation}
The fourth component of $V_\mu$ may be identified with the standard WF $\rho$. 
As a check on this quantity, its integral w.r.t. $d^4p$ yields the usual
4-current $J_\mu(x)$ as
\begin{equation}
J_\mu(x) = \int d^4p V_\mu(x,p) = {\bar \psi}(x)i\gamma_\mu \psi(x)
\end{equation}
from which the standard current conservation equation is recovered as
\begin{equation}
{\partial}_\mu J_\mu(x) = 0.
\end{equation}   
Next we make a Gordon reduction of $V_\mu$ (always possible for a free 
field $\psi(x)$) so as to bring out the separate contributions $V_\mu^c$ 
and $V_\mu^s$ arising from the `convective' and `spin' currents respectively:
\begin{equation}
V_\mu^c(x,p) = {\pi}^{-4} \int d^4q e^{2ip.q} 
{\bar \psi}(x+q) {{p_\mu} \over m} \psi(x-q) ; 
\end{equation}
\begin{equation}
V_\mu^s(x,p) = {\pi}^{-4} \partial_\nu^{(x)} \int d^4q e^{2ip.q} 
{\bar \psi}(x+q){{\sigma_{\mu \nu}} \over {2m}} \psi(x-q).
\end{equation}
We check from these that {\it two} separate conservation equations follow:
\begin{equation}
\partial_\mu V_\mu^c (x,p) = \partial_\mu V_\mu^s(x,p) = 0,
\end{equation}
the former being a consequence of the K-G equation for each component of
$\psi(x)$ and ${\bar \psi}(x)$, and the latter being identically satisfied.   
\par
	We now seek to use the Gordon reduction as a key element for deriving 
the Liouville equation a la [4], and in so doing bring out its role in 
putting our procedure in closer correspondence to [4]. For, the central issue 
is one of demonstrating that the density function $\rho(x,p)$ and the current 
function $V_i(x,p)$ are related by the `classical' velocity factor $v_i$, so
as to ensure the `correct' structure of the Liouville equation [4]. To 
that end we note from (2.3,6,7) that the effect of Gordon reduction in (3) 
may be expressed in operator form sandwiched between $\psi$ and ${\bar \psi}$:
\begin{equation}
i \gamma_\mu = {{p_\mu} \over {m}} + {1 \over {2m}}\partial^{(x)}_\nu 
\sigma_{\mu\nu}
\end{equation}   
The spatial part of this equation is seen as the precise counterpart of 
eq.(27) of [4] (after integration w.r.t. $d^4p d^4q$), with the following 
identifications: The convective term $p_\mu$ in (2.9) has a space part $p_i$ 
which corresponds to the $j_i$ term of [4], since it bears the ratio $v_i$
to the time component $p_0$, as the correct factor connecting the current 
function $V_i(x,p)$ to the density function $V_0(x,p)$ . The space part of 
the full $\gamma_\mu$ term collectively represents the two terms $a_i{\it d}$ 
and $a_i\epsilon$ in eq.(27) of [4]. Finally the last (derivative) term in 
(2.9) corresponds to the $ {\bf \nabla} \times {\bf \Phi} $ term in eq.(27) of 
[4] since it has a zero divergence by itself, as seen from the second part of 
eq.(2.8), and therefore has no separate influence on the equation of 
continuity. Therefore, following the logic of [4], this term has no effect on 
the anatomy of the Liouville equation which is thus entirely contained 
in the first (convective) part of eq.(2.8). To reveal this structure more 
explicitly, the $\rho(x,p)$ function of [4] reads in our $4 \times 4$ 
Dirac matrix notation, as follows: 
\begin{equation}
\rho(x,p) \equiv  V_0(x,p) = {\pi}^{-4} \int d^4q e^{2ip.q}  
{\bar \psi}(x+q)\gamma_4 \psi(x-q). 
\end{equation}
which satisfies the Liouville equation 
\begin{equation}
\partial_0 \rho(x,p) + v_i \partial_i \rho(x,p) = 0
\end{equation} 
that is merely a paraphrase of the first part of eq.(2.8) representing the 
conservation of the {\it convective} part of the $V_\mu$ function.   
\par
	To end this section, we outline a construction for the $\rho$ 
function analogously to eqs.(51-53) of [4], using a fully covariant procedure. 
To that end, we write down Fourier representations for the Dirac spinors 
on the lines of [12], but adapted to the Euclidean phase convention [13]:
\begin{equation}
\psi(x_1) = (2\pi)^{-3/2} \int d^3p_1 (m/E_1)^{1/2} [b_{p_1r_1} u^{r_1}(p_1)
e^{ix_1.p_1} + d^*_{p_1r_1} v^{r_1}(p_1) e^{-ix_1.p_1}] ;
\end{equation} 
\begin{equation}
{\bar \psi}(x_2) = (2\pi)^{-3/2} \int d^3p_2 (m/E_2)^{1/2} [b^*_{p_2r_2} 
{\bar u}^{r_2}(p_2) e^{-ix_2.p_2} + d_{p_2r_2} {\bar v}^{r_2}(p_2) \times 
e^{ix_2.p_2}] ;
\end{equation}
where $b$ and $d$ are (c-number) electron and positron amplitudes 
respectively and $u$ and $v$ are the corresponding positive and negative 
energy spinors. The latter are more conveniently expressed in terms of 
Lorentz boost operators as follows [12]:
\begin{equation}
u^{r_1}(p_1) = [(m-i\gamma.p_1)/ (2\sqrt{m(m+E_1)})] u^{r_1}(0); 
\end{equation}
\begin{equation}
v^{r_1}(p_1) =[(m+i\gamma.p_1)/ (2\sqrt{m(m+E_1)})] v^{r_1}(0).
\end{equation}
These Lorentz boost factors are proportional to energy projection operators
for the corresponding spinors. To specify their spin states, we can further
multiply these quantities by the spin projection operators $P_n$ [12] 
\begin{equation}
 P_n = [1 + i\gamma.{\hat n} \gamma_5]/2 ,
\end{equation}
where ${\hat n}$ is a unit 4-vector representing the direction of spin [12].
After these manipulations on the spinors through the sequence of equations 
(2.10-16), and their substitution in the defining eq.(2.3) for $V_\mu$, the 
spinor dependence can be totally removed by `tracing'. To simplify these 
expressions note that only the $b^*b$ and $d^*d$ terms will survive, while 
the cross terms will drop out of the traces, since the $\pm$ energy spinors
at a given momentum (zero) are orthogonal to each other. The rest of the 
simplification is routine and  the result can be compactly expressed in terms 
of two quantities $Tr^{\pm}_\mu$ defined as follows:
\begin{equation}
Tr^{\pm}_\mu = {{\pm m(p_1+p_2)_\mu + \epsilon_{\mu\lambda\rho\sigma}
p_{1\lambda} p_{2\rho}{\hat n}_\sigma} \over {2 \sqrt {E_1(m+E_1)E_2(m+E_2)}}}
\end{equation}
Before substituting these quantities in eq.(2.3) for $V_\mu(x,p)$, note that
of the two terms on the right, the first (convective) term has the desired
structure for the `Liouville form', eq.(2.11), but the second (spin) term 
has not. However, in accordance with the discussion immediately following 
eq.(2.9), in line with [4], one need not insist on the `Liouville connection' 
separately for both the terms since the `spin' term (proportional to 
${\hat n}$) can be traced back to the `derivative' term of (2.9) which has
no separate effect on the equation of continuity. Therefore we need only 
the $\mu = 4$ component in eq.(2.17) above to extract the $\rho$-function 
fully, whence the corresponding `current' function $V_i$ can be generated 
by multiplying with $v_i= p_i/p_0$ only.  
\par
	To proceed further, integration over $d^3p_1$ and $d^3p_2$ may be 
simplified via the transformations $p_{1,2} = p' \pm k/2$, remembering 
that $x_{2,1} = x \pm q$. Further, the 3D integrals may be converted to
4D integrations over $p'$ and $k$ via the identity
\begin{equation}
d^3p_1 d^3p_2 = 2E_12E_2 d^4p'd^4k \delta(p'^2+m^2+k^2/4) \delta(2p'.k)
\end{equation}
making use of the result $\delta(a)\delta(b) = \delta((a+b)/2)\delta(a-b)$.
Integration over $d^4q$ now gives $\delta^4(p'\mp p)$ for the terms $b^*b$ 
and $d^*d$ respectively; subsequent integration over $d^4p'$ gives 
$p'=\pm p$. Collecting all these results and substituting in (2.10), gives
\begin{eqnarray}
\rho(x,p) & = & {\pi}^{-3}\int d^4k \delta(p^2+m^2+k^2/4)\delta(2p.k) \nonumber \\
          &   & \times [b^*({\bf {p+k/2}})b({\bf {p-k/2}}) e^{ik.x} + 
d^*({\bf {p+k/2}})d({\bf {p-k/2}}) e^{-ik.x}] \nonumber \\  
          &   & \times (2mp_0 + 2i {\hat n}.({\bf p} \times {\bf k})) 
{\sqrt {{E_1E_2} \over {(m+E_1)(m+E_2)}}}       
\end{eqnarray}
This expression corresponds to eqs.(51-53) of [4], after taking due account 
of the difference in the choice of the respective basis functions (Dirac- 
vs. Pauli- types), as well as the covariant vs. non-covariant notations.
Note that the integration in (2.19) is basically angular $d\Omega_k$ in 
content, like in [4], since the 4D integration $d^4k$ is also multiplied by 
the two $\delta$-functions which take care of the magnitudes of both 
${\bf k}$ and $k_0$. Our 3-vector ${\bf k}$ is $2{\bf k}$ of [4], and 
${\hat n}$ is ${\hat \omega}$ of [4]. Finally, our expression for $\rho$ 
compactly covers both $\pm$ energy components.
\par
	We next turn to the K-G and Proca cases via Kemmer-Duffin equations.

\section{Wigner Function for Integral Spin}
\setcounter{equation}{0}
\renewcommand{\theequation}{3.\arabic{equation}}

	Because of the relative lack of exposure of the Kemmer-Duffin
equations [8] in the contemporary literature, it will first be useful to 
recall some basic results which are summarized in Roman [13] in the 
Euclidean phase convention, from which we shall freely draw on several 
results without much comment. The first order differential equation for 
integral spin is expressible in $\beta$- matrix notation as [13]
\begin{equation}
(\beta_\mu \partial_\mu + m) \psi(x) = 0; \quad  {\bar \psi}(m- \partial_\mu
\beta_\mu) = 0
\end{equation}
where $\psi(x)$ is a multicomponent field, each of whose components $\psi_i$
satisfies the K-G equation, and ${\bar \psi}=  \psi^{\dagger}\eta_4$ and
$\eta_4 = 2\beta_4^2 - 1$. The spin-0 and spin-1 cases correspond to 5- and
10- component irreducible representations respectively. These components are      
\begin{eqnarray}
Spin-0: \psi_i (i=1-4) & = & -\partial_\mu \phi(x)/m; \quad  \psi_5 
= \phi \nonumber \\
Spin-1: \psi_i (i=1-6) & = & -F_{\mu\nu}/m ; \quad  \psi_i (i=1-4) = \phi_\mu
\end{eqnarray}       
The various quantities like bilocal 4-current $j_\mu(x_1,x_2)$, phase
space 4-current $V_\mu(x,p)$, and local 4-current $J_\mu(x)$ follow very
closely the corresponding constructions (1.2-5) for the Dirac case, with
the replacements: $\gamma_\mu$ by $\beta_\mu$ ; $\gamma_4$ by $\eta_4$. 
Thus the phase space 4-current is given by
\begin{equation}
V_\mu(x,p) = {\pi}^{-4} \int d^4q e^{2ip.q} {\bar \psi}(x+q) i\beta_\mu
\psi(x-q)
\end{equation}
and its break-up into convective and spin components is achieved via
`Gordon reduction' analogously to the Dirac case. However the algebra of 
the $\beta$ matrices is a bit more involved than that of the $\gamma$ 
matrices because the former are singular. Thus [8,13]
\begin{equation}
\beta_\lambda \beta_\mu \beta_\nu + \beta_\nu \beta_\mu \beta_\lambda =
\beta_\lambda \delta_{\mu\nu} + \beta_\nu \delta_{\mu\lambda}; \quad
\eta_4^2 = 1
\end{equation} 
The `spin' operator $S_{\mu\nu}$ is given by 
\begin{equation}
iS_{\mu\nu} = \beta_\mu\beta_\nu - \beta_\nu\beta_\mu
\end{equation}
and, except for a factor of half, has exactly the same role as that of
$\sigma_{\mu\nu}$, including its antisymmetry property. For writing down the 
counterpart of Gordon reduction, the following equations are needed:
\begin{equation}
(\partial_\mu - \beta_\nu \beta_\mu \partial_\nu) \psi(x) = 0; \quad
{\bar \psi}(x) (\partial_\mu - \beta_\nu \beta_\mu \partial_\nu) = 0
\end{equation}
With the help of (3.1) and (3.6), and making use of the definition (3.5)
for the `spin', the matrix $\beta_\mu$ in (3.3) may be written in a form
analogously to (2.9) as
\begin{equation}
i\beta_\mu = {{p_\mu} \over m} + \partial^{(x)}_\nu {{S_{\mu\nu} \over m}}
\end{equation}  
which shows a clear division between the convective and spin parts, with
the latter having a `derivative' structure whose divergence vanishes by
itself, as in the second part of eq.(2.8). Therefore this term has no direct 
role in the Liouville equation like in the Dirac case, exactly as in [4].  
\par
	So far this $\beta$-formalism for integral spin covers both spin-0
(K-G) and spin-1 (Proca) cases, and one may wonder why for the former there
should be a spin term $S_{\mu\nu}$ at all, in the Gordon reduction (3.7).
Actually this term is effectively {\it zero} for the spin-0 case since, as 
has been explained in Roman [13], for the $5 \times 5$ representation the
eigenvalues of $S_3 = S_{12}$ are non-measurable. Indeed the probability
density [13]
\begin{equation}
\rho = -i(\psi^*_4 \psi_5 + \psi^*_5 \psi_4) = -i(\partial_0 \phi^* \phi
-\phi^* \partial_0 \phi)/m
\end{equation}    
vanishes in the eigenstates that correspond to $S_3 = \pm 1$, so that the
presence of the $S_{\mu\nu}$-term in the spin-0 case is purely symbolic.
Substituting this information in (3.3), and noting from (3.8) that only the 
components $\psi_{4,5}$ survive for the  phase space density function 
$\rho(x,p)$ for the K-G case, the final result works out as
\begin{equation}
\rho(x,p) = {1 \over m} {\pi}^{-4} i \int d^4q e^{2ip.q} 
[\partial_0 \phi^*(x+q) \phi(x-q) - \phi^*(x+q) \partial_0 \phi(x-q)]
\end{equation}     
while eq.(3.7) ensures that the phase space current function ${\bf V}(x,p)$
is related to $\rho(x,p)$ by the factor ${\bf p}/p_0 = {\bf v}$, so that
the structure of the Liouville equation is preserved. As a further check, the
integration of (3.9) over $d^4p$ reproduces the K-G density function 
$\rho(x)$ of (3.8) exactly. The last quantity is of course not positive 
definite, as expected [13] for the K-G case. Again as in the Dirac case,
it is possible to write down an expression for $\rho(x,p)$ closely analogous 
to (2.19), except for the absence of the `spin' term, in the third line. We
omit this routine step for brevity. 
\par
	For the Proca case, the `spin' term in the Gordon reduction (3.7)
will now contribute to the $\rho(x,p)$ function, while the Liouville 
connection is again maintained via the convective $p_\mu$ term only. From
the structure of $\rho(x,p)$, which corresponds to $\mu = 4$, it is seen 
quite clearly that the `spin' contribution  arising from the second term of 
eq.(3.7) in the Proca case is proportional to ${\bf S}.{\bf E} \times 
{\bf H}$, where ${\bf E}$ and ${\bf H}$ are the `electric' and `magnetic' 
components of $F_{\mu\nu}$ respectively. Again we skip the Fourier 
representation (2.19) for the full $\rho(x,p)$ function, as this step will 
not illuminate the structure any further beyond the above demonstration of 
a close parallelism between the spin- 1/2 and integral spin (0,1) cases.              

\section{Discussion and Conclusion}

This study was motivated by a desire to understand some results due to
Bosanac [4] in which he claims to derive the Dirac equation, starting 
from the (classical) premises of the Liouville equation. What intrigued
us was the assertion [4] that the `derivation' leads to the Dirac equation,
instead of to the Klein-Gordon equation. To examine this question more
closely, we were led to start from the {\it opposite} direction to [4], viz., 
a `direct' derivation of the Liouville equation through an explicit 
construction of the Wigner Functions [7] themselves, as in more orthodox 
approaches [5,6], with a view to throwing some light on this apparent 
asymmetry between the Dirac and the K-G cases. Rather surprisingly, our
result indicates a close parallelism between  both spin-1/2 and spin-(0,1) 
cases, brought about by the Kemmer-Duffin [8] formalism [13] for integral 
spin as a natural counterpart to the Dirac equation for spin-1/2. In both
cases, we have made essential use of the Gordon reduction into `convective' 
and `spin' contributions to the 4-current, wherein the convective part gives 
the desired `Liouville ratio' $ p_0 : {\bf p}$ between the density and 
current functions in phase space, but the spin part does not. However the 
latter has a derivative structure by virtue of which its 4-divergence 
vanishes identically, so as to yield a `spin' current conservation by itself 
[eq.(2.8)], and hence has no direct effect on the dynamics of the Liouville 
equation as such, which is entirely governed by the convective term only. 
This result seems to accord rather well with the corresponding result of 
Bosanac [4], viz., his eq.(27), despite a vast difference in the respective
procedures. In particular, since Gordon reduction in our derivation has
proved crucial for a neat identification of the origin of these two terms as 
convective and spin contributions respectively, it may be safely assumed that
the same mechanism underlies the derivation in [4] as well, irrespective of
the manner of the derivation.    
\par
	Without going into the philosophical aspects of the classical
foundations of Quantum Mechanics, it is pertinent to ask why only the Dirac
equation has  been amenable to the Bosanac derivation [4]. For, it will be 
rather naive to argue that no substantial quantum mechanical input has 
gone into the latter approach, since the very structure of the Wigner
Function as a bilinear in the (quantum mechanical) wave function is a 
tacit acknowledgement of the existence of the latter, with the Uncertainty
Principle as a built-in corollary of the same. Further, the structure of the
Dirac equation is implicit in the ansatz [4], eq.(16), in terms of Pauli
spinors $F_1, F_2$, with anticommutation relations (23-25) characteristic
of Dirac matrices in the Pauli 2-component form, all of which ingredients
have been crucial to the (indirect) derivation of the main eq.(27). Indeed
the entire exercise [4] is a fine example of how to work backwards from the 
(bilinear) Liouville equation to the (linear) Dirac equation, albeit in the 
Pauli 2-component form. Yet it goes without saying that the ansatz of a 
4-component spinor (satisfying a Dirac-type equation (16) [4]) has been very 
much of an input in the derivation.          
\par
	A comparison with our `direct' derivation would suggest that a possible 
reason why the K-G equation was not amenable to the Bosanac derivation was
the missing element of a multicomponent wave function in his formulation of
the K-G problem, something we have sought to rectify through the Kemmer-
Duffin $\beta$-matrix formalism [8] which restores the formal analogy to the
Dirac case. However, and this may be important, even if the Bosanac formalism 
had included a multicomponent wave function for integral (including zero) 
spin field, it is doubtful if working backwards from the Liouville to the 
K-G equation would have proved successful in this case, since the $\beta$-
matrices are {\it singular}, unlike the Dirac matrices, so that a routine 
inversion of steps on the lines of [4] might well not have been possible; (we 
did not succeed in this effort). But there seems to be no problem in the
`forward' steps (from QM equation to Liouville) going through in the K-G 
and Proca cases with equal ease, as the above derivation indicates. 
\par
	The foregoing comparison would suggest that the structure of Quantum 
Mechanics is probably richer and more varied than can be fully captured in 
one go, through alternative formulations. And yet the basically indeterministic
nature of QM poses a continuous challenge to innovative ideas  designed to 
capture its full flavour from `objective-realist' premises, so as to provide a 
more `classical' approach to its uncanny predictions. Ref.[4] may be regarded 
as one more effort in this direction, as part of an ongoing movement [14] for 
resolving the foundational paradoxes of orthodox QM which has acquired fresh 
momentum with heavyweights for and against the Copenhagen Interpretation 
arrayed in comparable strengths on both sides of the Great Divide [15].

\end{document}